\documentstyle[12pt]{article}

\catcode`@=11
\def\dif{{\rm d}}
\def\deriv{\@ifnextchar[{\@deriv}{\@deriv[]}}
   \def\@deriv[#1]#2#3{\mathchoice%
{{\dif^{#1}#2\over\dif{#3}^{#1}}}{{\dif^{#1}#2/\dif{#3}^{#1}}}%
{{\dif^{#1}#2\over\dif{#3}^{#1}}}{{\dif^{#1}#2/\dif{#3}^{#1}}}}

\def\presup#1{{}^{#1}\kern-.15em\relax}      
\def\presub#1{{}_{#1}\kern-.12em\relax}      

\def\secteqno{\@addtoreset{equation}{section}%
\def\theequation{\thesection.\arabic{equation}}}
\def\endsecteqno{\def\theequation{\@ifundefined{chapter}%
{\arabic{equation}}{\thechapter.\arabic{equation}}}}
\newcounter{subequation}
\def\thesubequation{\alph{subequation}}
\def\sneqnarray{\stepcounter{equation}\let\@currentlabel=\theequation
\setcounter{subequation}{1}
\def\@eqnnum{{\rm (\theequation\thesubequation)}}
\global\@eqcnt\z@\tabskip\@centering\let\\=\@eqncr\let\@@eqncr=\@@sneqncr
$$\halign to \displaywidth\bgroup\@eqnsel\hskip\@centering
 $\displaystyle\tabskip\z@{##}$&\global\@eqcnt\@ne
 \hskip 2\arraycolsep \hfil${##}$\hfil
 &\global\@eqcnt\tw@ \hskip 2\arraycolsep $\displaystyle\tabskip\z@{##}$\hfil
  \tabskip\@centering&\llap{##}\tabskip\z@\cr}
\def\endsneqnarray{\@@sneqncr\egroup $$\global\@ignoretrue}
\def\@@sneqncr{\let\@tempa\relax
   \ifcase\@eqcnt \def\@tempa{& & &}\or \def\@tempa{& &}
   \else \def\@tempa{&}\fi
     \@tempa \if@eqnsw\@eqnnum\stepcounter{subequation}\fi
     \global\@eqnswtrue\global\@eqcnt\z@\cr}
\def\nobiblabels{\def\@lbibitem[##1]##2{\@bibitem{##2}}}
\catcode`@=12

\secteqno
\begin{document}


\title{{\bf More nonperturbative corrections to the fine
      \\ and hyperfine splitting in the heavy quarkonium}}

\author{{\sc A.\,Pineda}\\
        \small{\it{Departament d'Estructura i Constituents
               de la Mat\`eria}}\\
        \small{\it{and}}\\
        \small{\it{Institut de F\'\i sica d'Altes Energies}}\\
        \small{\it{Universitat de Barcelona}}\\
        \small{\it{Diagonal, 647}}\\
        \small{\it{E-08028 Barcelona, Catalonia, Spain.}}\\
        {\it e-mail:} \small{pineda@ecm.ub.es} }

\date{\today}

\maketitle

\thispagestyle{empty}

\begin{abstract}
The leading
nonperturbative effects to the fine and hyperfine splitting were
calculated some time ago. Recently, they have been used in order to
obtain realistic
numerical results for the lower levels in bottomonium systems. We point
out that
a contribution of the same order $O(\Lambda_{QCD}^4/m^3 \alpha_s^2)$ has
been overlooked. We calculate it in this paper.

\end{abstract}

\vfill
\vbox{
\hfill May 1996\null\par
\hfill UB-ECM-PF 96/11}\null\par

\clearpage



\section{Introduction}
\indent

\bigskip

The leading non-perturbative (NP) corrections to the fine and hyperfine
splitting were
calculated a few years ago with the theory developed by Voloshin and
Leutwyler \cite{Vol1,Leut} for heavy quarkonia.
Leutwyler already computed the hyperfine splitting for $n=1$, $l=0$ in
ref. \cite{Leut}. However, a contribution of the same order
$O(\Lambda^4_{QCD}/m^3\alpha^2_s)$ was overlooked. Later on,
a number of authors (Voloshin, Di Giacomo et al. and
Dosch and collaborators) calculated the fine and
hyperfine
splittings for different quantum numbers ($n=2,1; l=0,1$) in refs.
\cite{Dosch} where the above overlooked contribution was pointed out
though finally neglected.
\medskip

Recently, new efforts have been made to obtain realistic
numerical results from rigorous QCD in refs. \cite{Ynd1,Ynd2} where a
coherent picture of bottomonium with $n=1,2$ is provided.
Relativistic, radiative and
the leading NP corrections were put together for the first time.
Unfortunately, the above mention contribution was also omitted.
\medskip

The aim of this paper is twofold: (i) to complete the calculation up to
order $O( \Lambda^4_{QCD}/m^3\alpha^2_s)$ and (ii) to control the error
made by
neglecting this contribution. The last point is important if we
eventually would like
to improve these results by taking into account further orders in the
perturbative and NP expansion.
We also provide results for a wide range of quantum numbers.
\medskip

Let us briefly discuss how this contribution arises.
Consider the Breit-Fermi interaction in the octet potential.
\begin{equation}
V_8^{Coul} \rightarrow V_8^{Coul}+V_8^{BF}
\end{equation}
then the octet propagator changes into
\begin{equation}
{1 \over E-H_8} \rightarrow {1 \over E-H_8-V_8^{BF}} \simeq
{1 \over E-H_8} + {1 \over E-H_8} V_8^{BF} {1 \over E-H_8}
\end{equation}
where $H_8$ is the octet Coulomb Hamiltonian. This leads to the
following correction to the energy
\begin{equation}
\label{debf}
\delta E_8 = \,
- {\pi \langle \alpha_s G^2 \rangle \over 6N_c}
\langle (n, j, l, s) \vert
{\vec r} {1 \over E_n - H_8}  V_8^{BF} {1 \over E_n -
H_8} {\vec r} \vert (n, j, l, s) \rangle\,.
\end{equation}
$E_n$ is the Coulomb singlet bound state energy. $n$ is the principal quantum
number. $l$, $s$ and $j$ are the angular momentum, spin and total angular momentum, respectively.

Let us comment upon the relative size of this contribution.
Notice that it is $1/N_c$ suppressed (in
fact ${1 \over N^2_c-1}$) by the ratio of octet O(${-1 \over 2N_c}$) to
singlet O(${N^2_c-1 \over 2N_c}$)
coupling constant and also it is suppressed by large energy differences
between octet and singlet
states (recall that the octet potential is repulsive). From this
we can conclude that matrix
elements with a larger number of octet propagators are
numerically suppressed.

Notice also that $V_8^{BF}$ only affects the leading NP results for
the fine and hyperfine splitting. For the standard NP results found by
Leutwyler and Voloshin in ref. \cite{Vol1,Leut} (\ref{debf}) gives rise
to a subleading contribution which will be neglected in the following.
Therefore,
we only take into account the piece of $V_8^{BF}$ which contributes to
the fine and hyperfine splitting
\begin{equation}
V_8^{BF} \doteq V_8^{F}+V_8^{HF}= V_8^{LS}+V_8^{T}+V_8^{HF}
\end{equation}
where

\begin{equation}
V^{F}_8 = V^{LS}_8+ V^{T}_8 =
{-1 \over 2N_c} {3 \alpha_s \over 2m^2} {1 \over r^3} {\vec L}.
{\vec S} +
{-1 \over 2N_c} {\alpha_s \over 4m^2} {1 \over r^3} S_{12}( \hat
r)\,, \end{equation}
\begin{equation}
S_{12}( \hat r) = 2(3 S_1 \cdot \hat r S_2 \cdot \hat r - S_1 \cdot S_2)
\,, \end{equation}
\begin{equation}
 V_8^{HF} = {-1 \over 2N_c} {4\pi \alpha_s \over 3m^2} \delta ^3
(\vec r) {\vec S}^2 \end{equation}
and
\begin{equation}
\delta E_8 \doteq \delta E_8^{F}+\delta E_8^{HF}= \delta E_8^{LS}+\delta E_8^{T}+\delta
E_8^{HF}\,.
\end{equation}
We shall give analytical formulas for the above energies when available.

Let us define the physical fine and hyperfine splitting for $n=2$,
$l=1$. We use spectroscopic notation $2^{2s+1}P_j$. The generalization
to arbitrary quantum
numbers is straightforward. The fine splittings read
\begin{equation}
\Delta_{21} = M(2^3P_2)-M(2^3P_1)\,, \quad
\Delta_{10} = M(2^3P_1)- M(2^3P_0)
\end{equation}
while the hyperfine splitting reads
\begin{equation}
\Delta_{HF}
\equiv {\bar M}(2^3P)-M(2^1P_1)
\end{equation}
where
\begin{equation}
{\bar M}(2^3P) \equiv
{1 \over 9} (5 M(2^3P_2)+3M(2^3P_1)+M(2^3P_0))
\end{equation}
Typically, $\delta E^{HF}$ only contributes to $\Delta_{HF}$ and $\delta
E^F$ only does to $\Delta_{21}$ and $\Delta_{10}$. Certainly, in
our case $\delta
E^{HF}_8$ only contributes to $\Delta_{HF}$ and $\Delta_{21}$,
$\Delta_{10}$ only get contributions from $\delta E^{F}_8$ but
rather peculiarly, as we will see below, $\Delta_{HF}$ receives
contributions from $\delta E^{F}_8$,
 that is, $\overline {\delta E^F_8} \not= 0 $ where
\begin{eqnarray}
&& \overline {\delta E^F_8} =
\\
\nonumber
&&
{1 \over 6\,l +3}
\left\{(2\,l+3)\delta E^F_8(j=l+1) +(2\,l+1)\delta E^F_8(j=l)+
(2\,l-1)\delta E^F_8(j=l-1) \right\}
\end{eqnarray}
for a general $l$.
This is quite unusual and did not happen with the NP corrections
calculated previously in the literature where $\overline{ \delta E^F}=0$.

\bigskip

\section{Fine Splitting}
\indent

For the fine splitting we obtain
\begin{equation}
\delta E_8^{LS} =
{1 \over 2N_c} {3 \alpha_s \over 2m^2}
{\pi \langle \alpha_s G^2 \rangle \over 6N_c}
\sum_{l_{1},l_{2}=0}^{\infty} F(n,l;l_{1},l_{2}) G^{LS}
(j,m,l;l_{1},l_{2}) \,,
\end{equation}

\begin{equation}
\delta E_8^{T} =
{1 \over 2N_c} {\alpha_s \over 4m^2}
{\pi \langle \alpha_s G^2 \rangle \over 6N_c}
\sum_{l_{1},l_{2}=0}^{\infty} F(n,l;l_{1},l_{2}) G^{T}
(j,m,l;l_{1},l_{2}) \rangle\,.
\end{equation}

We have split the matrix elements in radial and angular integrals.
\begin{equation}
\label{F}
F(n,l,l_{1},l_{2})=
\langle R_{nl} \vert
r {1 \over E_n - H_8^{(l_1)}} {1 \over r^3}
{1 \over E_n - H_8^{(l_2)}} r \vert R_{nl} \rangle
\end{equation}
and
\begin{equation}
G^{LS} (j,l;l_{1},l_{2}) = \sum_{j_{1}=|l_{1}-1|}^{l_{1}+1} \sum_{j_{2}=|l_{2}-1|}^{l_{2}+1}
\langle (j, l, s=1) \vert
{\hat {r_i}} I_{j_{1},l_{1}} {\vec L}.{\vec S} I_{j_{2},l_{2}}
{\hat {r_i}}
\vert (j, l, s=1) \rangle
\end{equation}
\begin{equation}
G^{T} (j,l;l_{1},l_{2}) = \sum_{j_{1}=|l_{1}-1|}^{l_{1}+1} \sum_{j_{2}=|l_{2}-1|}^{l_{2}+1}
\langle (j, l, s=1) \vert
{\hat {r_i}} I_{j_{1},l_{1}} S_{12}(\hat r) I_{j_{2},l_{2}}
{\hat {r_i}}
\vert (j, l, s=1) \rangle
\end{equation}
where $R_{nl}(r)$ is the radial Coulomb wave function,
\begin{equation}
\vert (n, j, l, s) \rangle = \vert R_{nl} \rangle \vert (j, l, s)
\rangle\,, \end{equation}
\begin{equation}
{1 \over E_n - H_8} \vert l \rangle =
{1 \over E_n - H_8^{(l)}} \vert l \rangle\,,
\end{equation}
and $I_{j,l}$ is the identity in the subspace with total angular
momentum $j$, orbital momentum $l$ and $s=1$ (otherwise the matrix
element is $0$).

Eq. (\ref{F}) can be computed using the techniques shown in ref.
\cite{Vol2} but
we have not succeeded in finding a close analytical expression for
$F(n,l;l_{1},l_{2})$ although we did succeed for the angular
momentum functions. They read

\begin{eqnarray}
\nonumber
&& G^{LS} (j,l;l_{1},l_{2}) =
(-1)^{l+l_{1}+1} \delta_{l_{1},l_{2}} (2l+1) C(l,1,l_{1};00)^2
\\
&&
\times
\sum_{j_{1}=|l_{1}-1|}^{l_{1}+1}
(2j_{1}+1) \left\{
\matrix{
l_{1} &j_{1} & 1 \cr
j  &l  & 1 \cr}
\right\}^2
\left( {j_{1}(j_{1}+1)-l_{1}(l_{1}+1)-2 \over 2} \right)
\,,
\end{eqnarray}

\begin{eqnarray}
\nonumber
&&G^{T} (j,l;l_{1},l_{2}) =
(2l+1) C(l,1,l_{1};00) C(l,1,l_{2};00)
\\
&&
\times
\sum_{j_{1}=|l_{1}-1|}^{l_{1}+1}
(2j_{1}+1) \left\{
\matrix{
l_{1} &j_{1} & 1 \cr
j  &l  & 1 \cr}
\right\}
\left\{
\matrix{
l_{2} &j_{1} & 1 \cr
j  &l  & 1 \cr}
\right\}
\\
\nonumber
&&
\times
2 \left( \delta_{l_{1},l_{2}}-3C(j_{1},1,l_{1};00)C(j_{1},1,l_{2};00) \right)
\,.
\end{eqnarray}

We display the explicit expressions in two tables.

Finally, the final expressions for the energy corrections read

\begin{equation}
\label{dels}
\delta E^{LS}_8 = \delta_{s,1} {\alpha_s  \over {\tilde \alpha}_s }
{\pi \langle \alpha_s G^2 \rangle \over m^3 (C_F {\tilde \alpha}_s)^2}
fls[n,l,j]\,,
\end{equation}

\begin{equation}
\label{det}
\delta E^{T}_8 = \delta_{s,1} {\alpha_s \over {\tilde \alpha}_s}
{\pi \langle \alpha_s G^2 \rangle \over m^3 (C_F {\tilde \alpha}_s)^2}
ft[n,l,j]\,.
\end{equation}
${\tilde \alpha}_s$ was defined in ref. \cite{Ynd1}.

We write (\ref{dels}) and (\ref{det}) in this way in order to make
comparison with \cite{Ynd2} simpler.
We give some numbers for lower values of $n$.
\begin{center}
$$
(fls+ft)[1,0,1] = {-9868 \over 8128125} \,, \quad
(fls+ft)[2,0,1] = {-236464 \over 19780605} \,, \quad
$$
$$
(fls+ft)[2,1,2] = {15475732 \over 415392705} \,, \quad
(fls+ft)[3,1,2] = {{8134524806}\over {53682451515}} \,,
$$
$$
(fls+ft)[2,1,1] = {-452188 \over 19780605} \,, \quad
(fls+ft)[3,1,1] = -{{6745478}\over {59449005}}\,,
$$
\begin{equation}
(fls+ft)[2,1,0] = {-1235984 \over 11868363} \,, \quad
(fls+ft)[3,1,0] = -{{83789896}\over {219112047}}\,.
\end{equation}
\end{center}

In principle our contributions for
the fine splitting are quite small in comparison with the NP and
radiative
corrections coming from the wave functions. Nevertheless, the authors of
ref. \cite{Ynd2}
managed to isolate the latter in an unknown factor which is
determined from data. Hence, the remaining perturbative and NP
 effects are kept under control.
For the fine splitting
they obtain
\begin{equation}
\label{213}
\Delta_{10}={5 \over 4}(1+ \delta_{rad})\Delta_{21}- \delta_{NP}\,.
\end{equation}
They take $\Delta_{21}$ from
data and $\delta_{rad}$ and $\delta_{NP}$ are respectively the remaining
radiative and NP corrections (see (3.2) in the second paper
of \cite{Ynd2} for details). While, the new contribution reads (to be
added to $\delta_{NP}$ in (\ref{213}))

\begin{equation}
\delta_{NP}(new) =
{-5 \over 4} \delta E^F_8(j=2)+{9 \over 4} \delta E^F_8(j=1)- \delta
E^F_8(j=0) = - {\alpha_s \over {\tilde \alpha}_s}
{\pi \langle \alpha_s G^2 \rangle \over m^3 (C_F {\tilde \alpha}_s)^2}
{2548892 \over 415392705}
\end{equation}
being around $1\%$ smaller.

\section{Hyperfine Splitting}
\indent

In this section we work out the analytical formula for
the hyperfine splitting. It receives contributions from both
$\overline{ \delta E^F_8}$ and $\delta E^{HF}_8$. Let us calculate
$\delta E^{HF}_8$.
The angular integral is almost trivial. The delta function only survives
for $0$ angular momentum which after composing with ${\vec r}$
implies that only $l=1$ states contribute.
It is quite remarkable that only for $l=1$ states we obtain a non-zero
contribution. After these manipulations we find that $\delta E^{HF}_8$
becomes
\begin{equation}
\delta E^{HF}_8 = s(s+1) \delta_{l,1} {1 \over 2N_c} {\alpha_s \over 9m^2}
{\pi \langle \alpha_s G^2 \rangle \over 6N_c}
|
\langle R_{nl} \vert
r {1 \over E_n - H_8^{(l_1=0)}}
\vert r = 0 \rangle |^2\,.
\end{equation}
The radial integral can be carried out by following the formulas given
in \cite{Vol2}. We obtain
\begin{equation}
\label{dehf}
\delta E^{HF}_8 = \delta_{l,1} s(s+1) \alpha_s
{\pi \langle \alpha_s G^2 \rangle \over (m {\tilde \alpha}_s)^3}
hf[n]\,,
\end{equation}
\begin{equation}
hf[n]:= {n^5(n^2-1) \over 2^9}{\Gamma [9n/8-2]^2 \over
\Gamma[9n/8+3]^2}
\end{equation}

We again write (\ref{dehf}) in this way in order to make comparison
with \cite{Ynd2} simpler.
We give some numbers for lower values of $n$ ($n \geq 2$).

$$
hf[2] = {65536 \over 32967675} \,, \quad
hf[3] = {16777216 \over 296815671075} \,,
$$
\begin{equation}
hf[4] = {2048 \over 135270135} \,, \quad
hf[5] = {2097152000 \over 324918632936187} \,.
\end{equation}

Let us know discuss our contribution to $\Delta_{HF}$.
If $l=0$ it turns out to be quite small but for $l
\not= 0$ the leading perturbative order is 0. Therefore, the next
perturbative order is needed and in principle the nonperturbative
effects are going to be more important. Thus, in the second paper of
\cite{Ynd2} a careful analysis of the hyperfine splitting
was done for
$n=2$,
 $l=1$. It was obtained

\begin{equation}
\label{35}
\Delta_{HF}={5 \over 24}( {\beta_0 \over 2} -{21 \over 4})C_F \alpha_s
\Delta_{21}+
{976 \over 1053}
{\pi \langle \alpha_s G^2 \rangle \over m^3 (C_F {\tilde \alpha}_s)^2}
\,.
\end{equation}
While our contribution reads (to be added to (\ref{35}))

\begin{equation}
\Delta_{HF}(new)= \overline {\delta E^F_8} + \delta E^{HF}_8
= {\alpha_s \over {\tilde \alpha}_s}
{\pi \langle \alpha_s G^2 \rangle \over m^3 (C_F {\tilde \alpha}_s)^2}
{160277456 \over 18692671725}
\end{equation}
which turns out to be around $1 \%$ smaller.
\bigskip

\section{Discussion}
\indent

From the phenomenological point of view our results are going to be
important only in the event that a very high precision
 measurement is done.
 Nevertheless, they are conceptually
important since they take into account the
perturbative corrections
to the octet Coulomb potential. We have also seen that the hyperfine
splitting gets corrections from the fine octet potential which is
something somewhat unusual.

Our results complete the calculation of all the contributions of order
$O(\Lambda^4_{QCD}/m^3\alpha_s^2)$. We have also seen the error
in neglecting this contribution. This step is unavoidable for an
eventual improvement of these results by calculating higher orders in
the perturbative and NP expansion.
\bigskip

{\bf Acknowledgments}
\medskip

I acknowledge a fellowship from CIRIT.
 Financial support from CICYT, contract AEN95-0590 and financial
support from CIRIT, contract GRQ93-1047 is also acknowledged.
I thank J. Soto for the critical reading of the manuscript and
illuminating conversations.

\bigskip

\vfill
\eject

\begin{table}
\begin{center}
\begin{tabular}{|c|l|l|l|}  \hline
$G^{LS}$  & $j=l-1$ & $j=l$  & $j=l+1$ \\ \hline
$l_{1}= l-1$ & ${{1 - {l^2}}\over {1 + 2\,l}}$ & ${{1 - l}\over
{1 + 2\,l}}$   &
 ${{\left( -1 + l \right) \,l}\over {1 + 2\,l}}$  \\ \hline
$l_{1}=l$    & $0$     & $0$    & $0$   \\ \hline
$l_{1}=l+1$  & $-{{\left( 1 + l \right) \,\left( 2 + l \right) }\over {1
+ 2\,l}}$ &
 $-{{2 + l}\over {1 + 2\,l}}$ & ${{l\,\left( 2 + l \right) }\over {1 +
2\,l}}$
   \\ \hline
\end{tabular}
\end{center}
\caption{We display here $G^{LS}$. For $l=0$ $j$ must be $1$. The
remaining matrix elements are zero.}
\end{table}

\begin{table}
\begin{center}
\begin{tabular} {|c|l|l|l|l|} \hline
$G^T$ & $l_{1}=l-1, l_{2}=l-1$ & $l_{1}=l-1, l_{2}=l+1$ & $l_{1}=l+1,
l_{2} =l+1$ \\ \hline
$j=l-1$
& ${{2\,\left( -6 - 11\,l - 26\,{l^2} - {l^3} + 10\,{l^4} - 12\,{l^5} -
        8\,{l^6} \right)}\over
    { l\,\left( -3 + 2\,l \right) \,
   \left( -1 + 2\,l \right) \,{{\left( 1 + 2\,l \right) }^3}}}^{(*)}$ &
$ {{-6\,\left( 1 + l \right) }\over {{{\left( 1 + 2\,l \right) }^3}}}$ &
$ {{2\,\left( 1 - 2\,l \right) \,\left( 1 + l \right) \,
      \left( 2 + l \right) }\over {{{\left( 1 + 2\,l \right) }^3}}}$
\\ \hline
$j= l$ & ${{2\,\left( -1 + l \right) \,
      \left( 2 + 3\,l + 8\,{l^2} + 4\,{l^3} \right) }\over
    {l\,\left( -1 + 2\,l \right) \,{{\left( 1 + 2\,l \right) }^2}}}$ &
${6\over {{{\left( 1 + 2\,l \right) }^2}}}$ &
$  {{2\,\left( -6 - 5\,l + 7\,{l^2} + 12\,{l^3} + 4\,{l^4} \right)
}\over
    {{{\left( 1 + 2\,l \right) }^2}\,\left(3+5\,l+ 2\,{l^2} \right)}}$
\\ \hline
$ j= l+1$ & ${{2\,\left( 1 - l \right) \,l\,\left( 3 + 2\,l \right)
}\over
    {{{\left( 1 + 2\,l \right) }^3}}}$ &
$  {{-6\,l}\over {{{\left( 1 + 2\,l \right) }^3}}} $&
$  {{2\,\left( -6 + 14\,l + 37\,{l^2} + {l^3} - 50\,{l^4} - 36\,{l^5} -
        8\,{l^6} \right) }\over
    {\left( 1 + l \right) \,{{\left( 1 + 2\,l \right) }^3}\,
      \left( 3 + 2\,l \right) \,\left( 5 + 2\,l \right) }}$ \\ \hline
\end{tabular}
\end{center}
\caption{We display here $G^T$. For $l=0$ $j$ must be $1$. The remaining
matrix elements are either
zero or they can be deduced by exchanging $l_{1} \leftrightarrow l_{2}$.
The asterisk indicates that the result is only valid for $l \geq 2$
otherwise it is $0$).} \end{table}



\begin{thebibliography}{99}

\bibitem{Vol1} M. B. Voloshin,
{\it Nucl. Phys.} {\bf B154} (1979) 365.

\bibitem{Leut} H. Leutwyler,
{\it Phys. Lett.} {\bf B98} (1981) 447.

\bibitem{Dosch} A. Kr\"amer, H. G. Dosch and R. A. Bertlmann, {\it
Phys. Lett.} {\bf B223} (1989) 105; {\it Fort.
der Phys.} {\bf 40} (1992) 93.
Voloshin, {\it Sov. J. Nucl. Phys.} {\bf 35} (1982) 592.
M. Campostrini, A. Di Giacomo, S. Olejnik, {\it Z. Phys.} {\bf C31}
(1986) 577. G. Curci, A. Di Giacomo, G. Paffuti {\it Z. Phys.} {\bf C18}
(1983) 135.

\bibitem{Ynd1} S. Titard and F.J. Yndur\'ain,
{\it Phys. Rev.}
{\bf D49} (1994) 6007.

\bibitem{Ynd2} S. Titard and F.J. Yndur\'ain,
{\it Phys.
Rev.} {\bf D51} (1995) 6348 and {\it Phys. Lett.} {\bf B351} (1995) 541.

\bibitem{Vol2} M. B. Voloshin,
Sov. J. Nucl. Phys., {\bf36}, 143 (1982).


\end{thebibliography}
\end{document}